\documentclass
[showpacs,onecolumn,12pt,aps,pra,superscriptaddress,floatfix]
{revtex4-1}
\usepackage{graphicx}% Include figure files
\usepackage{dcolumn}% Align table columns on decimal point
\usepackage{bm}% bold math
\usepackage{hyperref}% add hypertext capabilities
\usepackage{amsmath} % split in equation
\usepackage{color}
\usepackage{geometry}
\begin{document}
\title{Squeezing enhanced atom-cavity interaction in coupled cavities with high dissipation rates}

\author{Yan Wang}
\affiliation{Department of Physics, Harbin Institute of Technology, Harbin, 150001, China}

\author{Jin-Lei Wu}
\affiliation{Department of Physics, Harbin Institute of Technology, Harbin, 150001, China}

\author{Jie Song} \email{E-mail: jsong@hit.edu.cn}
\affiliation{Department of Physics, Harbin Institute of Technology, Harbin, 150001, China}

\author{Yong-Yuan Jiang}
\affiliation{Department of Physics, Harbin Institute of Technology, Harbin, 150001, China}

\author{Zi-Jing Zhang}
\affiliation{Department of Physics, Harbin Institute of Technology, Harbin, 150001, China}

\author{Yan Xia}
\affiliation{Department of Physics, Fuzhou University, Fuzhou, 350002, China}
\date{\today}
\begin{abstract}
The realization of the strong coupling regime is requisite for implementing quantum information tasks. Here, a method for enhancing the atom-field coupling in highly dissipative coupled cavities is proposed. By introducing parametric squeezing into the primary cavity which is only virtually excited under specific parametric conditions, coupling enhancement between atom and auxiliary cavity is realized for proper squeezing parameters. This enables the system to be robust against large cavity decay and atomic spontaneous emission. The observation of vacuum Rabi oscillations show that the originally weak-coupled system can be enhanced into effective strong coupling regime.
 
\end{abstract}
\maketitle

\section{Introduction}
Cavity quantum electrodynamics (QED) studies the light-matter interactions between cavity photons and quantum emitters \cite{bib1}, such as Rydberg and neutral atoms \cite{bib2,bib3,bib4}, superconducting qubits \cite{bib5,bib6}, and semiconductor quantum dots (QDs) \cite{bib7,bib8,bib9}. Strong coupling regime, where atom-cavity coupling strength has to be comparable or larger than atomic spontaneous emission rate $\gamma$ and cavity decay rate $\kappa$ \cite{bib10,bib11}, is indispensable for experimentally investigating a manifold of quantum phenomena and implementing quantum information processing (QIP) \cite{bib12}. Such strong interaction often requires resonators with high quality ($Q$) factor  and small mode volume ($V$) simultaneously, which is still difficult to engineer in experiments. However, flexible configurations for the cavities shift the mutual constraint between high $Q$ and small $V$. It is demonstrated that by employing a coupled cavity configuration \cite{bib13}, the requirement for high $Q$ and small $V$ for one cavity can be removed \cite{bib14}, hence, effective strong coupling in highly dissipative cavity QED system is realized. On the other hand, considerable efforts have been devoted to enhance the atom-cavity coupling strength. Parametric squeezing of the cavity mode is demonstrated to be feasible for exponentially enhancing atom-cavity coupling as well as the cooperativity \cite{bib15,bib16,bib17,bib18}.

It is well known that in cavity QED system, the excitation of cavity field can be eliminated completely by confining and coupling two atoms to a single cavity \cite{bib13,bib19}. For atoms which are highly detuned from the field mode, excitation may transfer between them without populating the field mode \cite{bib20}. Based on the method proposed in \cite{bib15,bib17}, we have demonstrated an enhancement of the dipole-dipole interaction between two atoms trapped in an optical cavity \cite{bib21}. In contrast to the single cavity scheme, in this paper, we put forward a scheme for enhancing the atom-cavity coupling in coupled cavities with high dissipation rates via parametric squeezing. This scheme contains unique advantages of the coupled cavity configuration proposed in \cite{bib14}, in particular it allows the primary cavity coupled to quantum emitter to be highly dissipative (i.e., low $Q$). The main progress in current work is summarized as follows. By employing parametric squeezing of the primary cavity mode, effective atom-cavity coupling and  coupling between neighbouring cavities can be enhanced greatly by adjusting the squeezing parameter. Although effective strong coupling between atom and auxiliary cavity can be established in the absence of cavity mode squeezing, which has been verified in \cite{bib14}, the requirement of the auxiliary cavity possessing high $Q$ remains difficult to implement for some real system. In this paper, we demonstrate that our scheme exhibits strong robustness against larger rates of both cavity decay and atomic spontaneous emission compared to previous study \cite{bib14} in the presence of the squeezing of the primary cavity mode. This could be of significant utility for relaxing the restriction in cavity $Q$. By choosing proper parameters, vacuum Rabi oscillations are observed, manifesting the highly dissipative coupled cavities system in the effective strong coupling regime. In addition, diverse dynamics of the system are also demonstrated in the proposed scheme.

The remainder of the paper is structured as follows. In section \ref{sec2.1}, we describe the physical model of the system and give the Hamiltonian in the presence of squeezing. In section \ref{sec2.2}, we use the Bogoliubov squeezing transformation to diagonalize the Hamiltonian and derive the effective Hamiltonian describing the indirect atom-cavity interaction under large detuning condition. In section \ref{sec3}, we show how the scheme enables extremely strong decay of both cavity mode and atom. The properties of the system are investigated under various parametric conditions where rich dynamics are demonstrated. Finally, we briefly discuss the possible experimental implementations of the scheme and summarize our conclusions in section \ref{sec4}.

\section{Coupling Enhancement Induced by Parametrically Squeezing the Cavity Mode}
\subsection{Model}\label{sec2.1}
\begin{figure}
	\includegraphics[width=0.8\columnwidth]{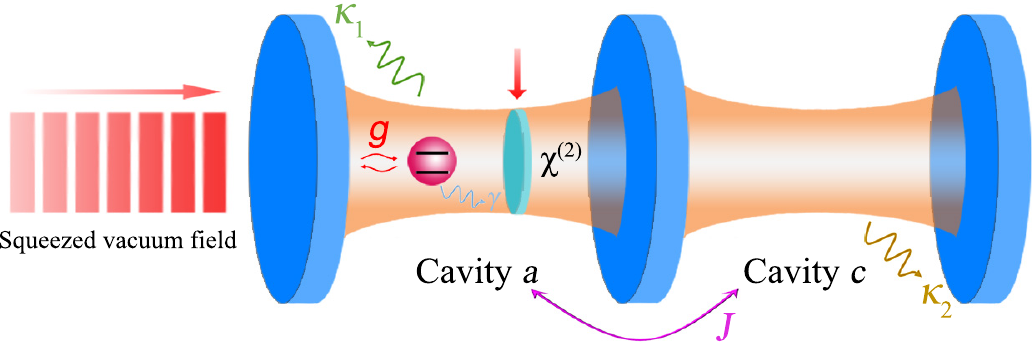}
	\caption{Schematic of the system. Two optical cavities are coupled with hopping rate $J$. A two level atom confined in the first primary cavity is coupled to cavity mode with coupling strength $g$. A $\chi^{(2)}$ nonlinear medium is used to induce a squeezed cavity mode, which is strongly pumped at frequency $\omega_p$, amplitude $\Omega_p$ and phase $\theta_p$. The primary cavity couples to a squeezed-vacuum reservoir that can be generated by optical parametric amplification with squeezing parameter $r_e$ and reference phase $\theta_e$. The decay rates of the atom and the two cavities are $\gamma$, $\kappa_1$, and $\kappa_2$, respectively.}
	\label{fig1}
\end{figure}

We consider a quantum system consisting of two coupled optical cavities, as sketched in Figure \ref{fig1}. The first primary cavity (with resonance frequency $\omega_a$ and decay rate $\kappa_1$) supports optical mode $a$, while the second auxiliary cavity (with resonance frequency $\omega_c$ and decay rate $\kappa_2$) supports optical mode $c$. Two cavities are coupled by the photon hopping with strength $J$, which can
be tuned by changing the distance between them. The primary cavity contains a two level atom with transition frequency $\omega_0$ and a $\chi^{(2)}$ nonlinear medium that is pumped with frequency $\omega_p$, amplitude $\Omega_p$ and phase $\theta_p$. Notably, the nonlinearity of the medium is used to induce a squeezed cavity mode \cite{bib17}. A high bandwidth squeezed vacuum field, which can be thought of as a squeezed vacuum reservoir (with squeezing parameter $r_e$ and reference phase $\theta_e$), is injected into the primary cavity \cite{bib15}. Experimentally, the proposed scheme could be implemented in the photonic crystal cavity, inspired by experimental advances in the coupled nanocavity arrays based on photonic crystals in present experiments \cite{bib1,bib8,bib9,bib22,bib23}. The squeezing environment is generally implemented via the process of optical parametric amplification (OPA) \cite{bib24,bib25}.

In the frame rotating at half the squeeze frequency $\omega_p/2$, the Hamiltonian of this system is given by ($\hbar$=1)
\begin{equation}\label{Eq:original H}
\begin{split}
H=\Delta_a a^\dag a+\Delta_c c^\dag c+\frac{\Delta_q}{2}\sigma_z+g(\sigma_+a+a^\dag\sigma_-)
+J(c^\dag a+a^\dag c)+\frac{\Omega_p}{2}(e^{\mathrm{i}\theta_p}a^2+e^{-\mathrm{i}\theta_p}{a^{\dag}}^2).
\end{split}
\end{equation}
Here, the two-level atom is described by the Pauli operator $\sigma_z=\vert e\rangle\langle e\vert-\vert g\rangle\langle g\vert$ and the transition operators $\sigma_+=\sigma_-^\dag=\vert e\rangle\langle g\vert$, where $\vert e\rangle$ and $\vert g\rangle$ are the excited state and the ground state, respectively. The detunings are $\Delta_a=\omega_a-\omega_p/2$, $\Delta_c=\omega_c-\omega_p/2$, and $\Delta_q=\omega_0-\omega_p/2$. 

\subsection{Enhancement of the Indirect Atom-cavity Coupling}\label{sec2.2}

The Hamiltonian can be diagonalized by introducing the Bogoliubov squeezing transformation $a=\cosh(r_p)a_s-e^{-\mathrm{i}\theta_p}\sinh(r_p)a_s^\dag$ \cite{bib21,bib26}, where the squeezing parameter $r_p$ is defined as  $r_p=(1/2)\mathrm{arctanh}(\Omega_p/\Delta_a)$, reads
\begin{equation}\label{Eq:squeezing frame H}
\begin{split}
H'=&\Delta_sa_s^\dag a_s+\Delta_cc^\dag c+\frac{\Delta_q}{2}\sigma_z+\frac{g}{2}[e^{r_p}(a_s^\dag+a_s)(\sigma_++\sigma_-)-e^{-r_p}(a_s^\dag-a_s)(\sigma_+-\sigma_-)]\\
&+\frac{J}{2}[e^{r_p}(a_s^\dag+a_s)(c^\dag+c)-e^{-r_p}(a_s^\dag-a_s)(c^\dag-c)],
\end{split}
\end{equation}
where $\Delta_s=\Delta_a\mathrm{sech}(2r_p)$ denotes the squeezed cavity frequency, and $\theta_p$ is set to zero for simplicity. Here, we assume that the cavity mode $a$ is initially in the vacuum state. Under the rotating-wave approximation (neglecting the terms that oscillate with high frequencies $\Delta_s+\Delta_q$, $\Delta_s+\Delta_c$) and the large detuning condition [$\Delta_s-\Delta_q\gg (g/2)e^{r_p}$, $\Delta_s-\Delta_c\gg (J/2)e^{r_p}$], the primary
cavity mode $a$ (with extremely high decay rate) can be eliminated adiabatically and we obtain the effective interaction between the atom and the auxiliary cavity mode $c$, with the effective Hamiltonian \cite{bib27}
\begin{equation}\label{Eq:effective H}
	H_{\mathrm{eff}}^1=\Delta_c'\,c^\dag c+\Delta_e\vert e\rangle\langle e\vert+g_{\mathrm{eff}}[\sigma_+c\,e^{\mathrm{i}(\Delta_q-\Delta_c)t}+\mathrm{H.c.}].
	\end{equation}
The effective detunings are $\Delta_c'=J^2\cosh^2({r_p})/(\Delta_c-\Delta_s)$ and $\Delta_e=g^2\cosh^2({r_p})/(\Delta_q-\Delta_s)$.
The third term in Eq. (\ref{Eq:effective H}) describes the enhanced atom-cavity interaction, with a controllable strength
\begin{equation}\label{Eq:effective coupling}
g_{\mathrm{eff}}=\frac{gJ\cosh^2({r_p})}{2}\Big(\frac{1}{\Delta_c-\Delta_s}+\frac{1}{\Delta_q-\Delta_s}\Big).
\end{equation}
In this case, the enhancement in atom-cavity coupling can be realized by simply adjusting the squeezing parameter $r_p$.
In addition, to realize the coherent energy exchange between the atom and the cavity field, the parametric condition for resonant interaction $\Delta_e-\Delta_c'+\Delta_q-\Delta_c=0$ should also be satisfied. 

\section{Robustness against strong dissipations}\label{sec3}
\begin{figure}
	\centering
	\includegraphics[width=0.8\columnwidth]{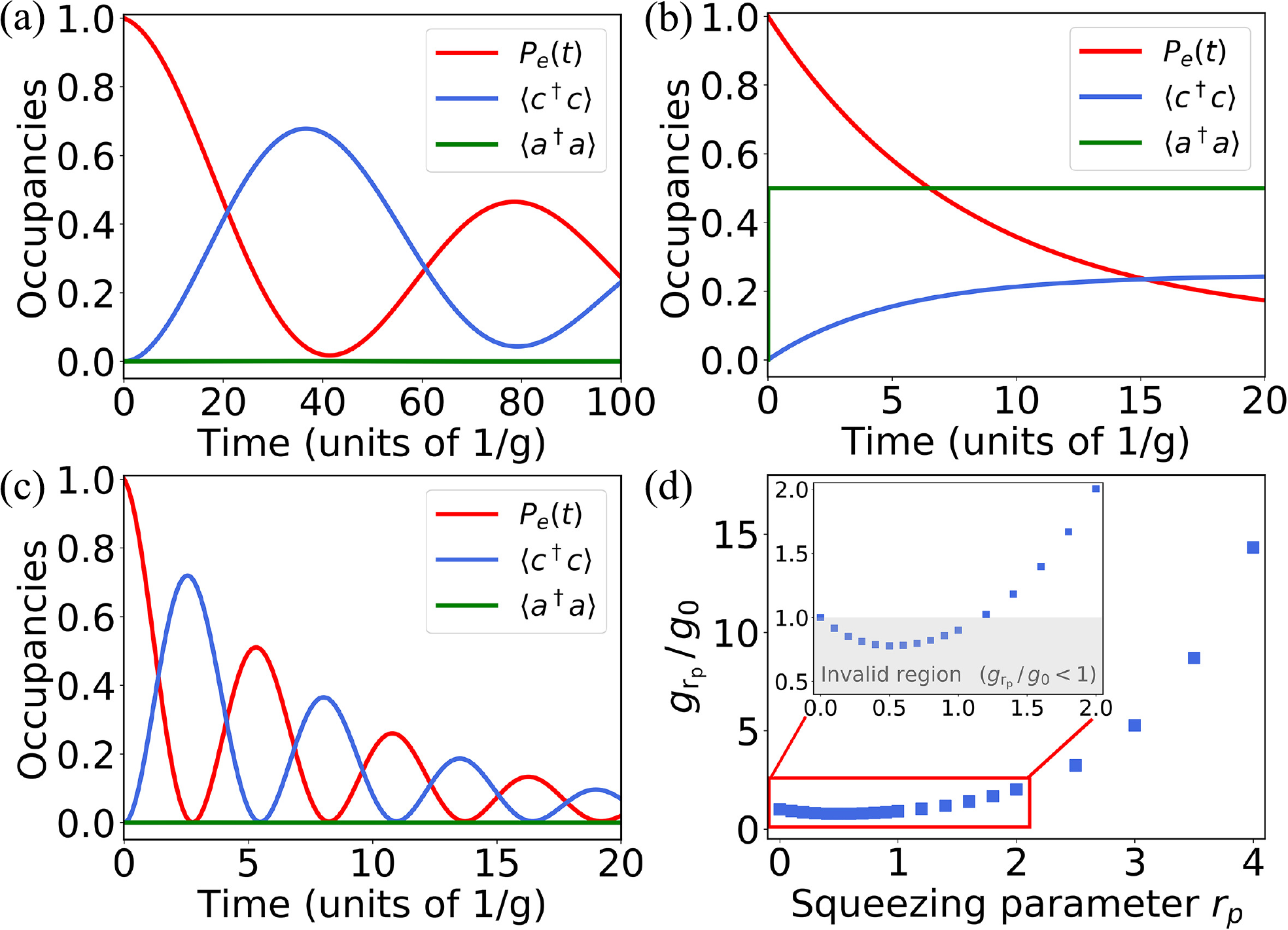}%
	\caption{(a-c) Time evolution of the mean photon numbers $\langle a^\dagger a\rangle$ (green curves), $\langle c^\dagger c\rangle$ (blue curves), and the probability of the atom being in the excited state $P_e(t)$ (red curves). In (a): $r_p=r_e=0$, $\theta_e=0$, $\kappa_1=10g$, and $\kappa_2=\gamma=10^{-3}g$. In (b): $r_p=4$, $r_e=0$, $\theta_e=0$. In (c): $r_p=r_e=4$, $\theta_e=\pi$. In (b),(c): $\kappa_1=100g$, and $\kappa_2=\gamma=0.1g$. In (a-c): $\theta_p=0$, $J=2g$, and the values of the detunings satisfy: $\Delta_s-\Delta_q=50ge^{r_p}$; $\Delta_s+\Delta_q=20(\Delta_s-\Delta_q)$; $\Delta_e-\Delta_c'+\Delta_q-\Delta_c=0$. The initial states of the two cavities are in the vacuum states, and the atom is in the excited state. (d) Enhancement of atom-cavity coupling versus squeezing parameter $r_p$. The ratio $g_{r_p}/g_0$ are obtained by comparing the oscillation periods for various $r_p$ with the oscillation period for $r_p=0$. The inset shows a clear view of the invalid region ($g_{\mathrm{r_p}}/g_0<1$).}
	\label{fig2}
\end{figure}

We now take the dissipations into consideration. In the absence of cavity mode squeezing, the dynamics of the system (as depicted in Figure \ref{fig1}) can be described by the master equation 
\begin{equation}\label{Eq: master equ before squeezing}
\begin{split}
\dot{\rho}(t)=&i[\rho(t),H(t)]+\gamma\mathcal{L}[\sigma_-]\rho(t)+\kappa_2\mathcal{L}[c]\rho(t)+\kappa_1(N+1)\mathcal{L}[a]\rho(t)+\kappa_1N\mathcal{L}[a^\dag)]\rho(t)\\
&-\kappa_1M\mathcal{L}'[a]\rho(t)-\kappa_1M^\ast\mathcal{L}'[a^\dag]\rho(t),
\end{split}
\end{equation}
where $\mathcal{L}(o)\rho=o\rho o^\dag-(o^\dag o\rho+\rho o^\dag o)/2$, $\mathcal{L}'(o)\rho=o\rho o-(oo\rho+\rho oo)/2$; $\kappa_1$, $\kappa_2$ and $\gamma$ are the decay rates of the primary cavity $a$, the auxiliary caivty $c$ and the atom, respectively; $N=\sinh^2(r_e)$ denotes the mean photon number of the squeezed field, and $M=\cosh(r_e)\sinh(r_e)e^{\mathrm{i}\theta_e}$ denotes the strength of the two-photon correlation \cite{bib28}. In Figure \ref{fig2}a, we plot the time evolution of the mean photon numbers $\langle a^\dagger a\rangle$, $\langle c^\dagger c\rangle$ and the probability of the atom being in the excited state $P_e(t)$, by numerically solving the master equation (\ref{Eq: master equ before squeezing}). Here, we first consider the case that the cavity mode $a$ is coupled to a thermal vacuum bath (i.e., $r_e$=0). The squeezed field is employed specifically for resisting the noise induced by cavity mode squeezing (see below). The primary cavity and the trapped atom are originally set as a weak-coupled cavity QED system with $\kappa_1\gg g$.
Typical vacuum Rabi oscillation is clearly observed, indicating the effective strong coupling between the atom and the cavity mode $c$. The extremely small occupancy of $\langle a^\dagger a\rangle$ confirms that the cavity mode $a$ is only virtually excited. Therefore, a large cavity decay rate $\kappa_1=10g$ can be chosen. This relaxes the restriction of high $Q$ for the resonators, especially the photonic crystal cavities possessing relatively low $Q$ factors. It is noteworthy that the decay rate of the auxiliary cavity is required to be low ($\kappa_2/g=10^{-3}$) for effective atom-cavity interaction in the present scenario. Nonetheless, the experimentally reported figure-of-merit $g/\kappa$ for a photonic-crystal nanocavity-QDs coupled system is limited to below 10 \cite{bib9}.

Perhaps more interesting is the ability of our scheme to allow larger decay rates of both atom and two cavities via parametrically squeezing the cavity mode. However, the squeezing of the cavity mode can introduce additional noises into the cavity, which immensely destroys the dynamics of the system (see Figure \ref{fig2}b). In principle, squeezing the cavity mode induces an enhancement in the system-reservoir coupling. In view of this, we employ the auxiliary squeezed field (reservoir) to offset the enhancement in system-reservoir coupling by appropriately parametric matching. This is equivalent to coupling the squeezed cavity mode to an effective vacuum reservoir. The method was proposed in \cite{bib15,bib17} and also generalized in our recent study \cite{bib21}. The master equation in terms of $a_s$ is re-expressed as
\begin{equation}\label{Eq: master equ after squeezing}
\begin{split}
\dot{\rho}(t)=&i[\rho(t),H'(t)]+\gamma\mathcal{L}[\sigma_-]\rho(t)+\kappa_2\mathcal{L}[c]\rho(t)+\kappa_1(N_s+1)\mathcal{L}[a_s]\rho(t)+\kappa_1N_s\mathcal{L}[a_s^\dag)]\rho(t)\\
&-\kappa_1M_s\mathcal{L}'[a_s]\rho(t)-\kappa_1M_s^\ast\mathcal{L}'[a_s^\dag]\rho(t),
\end{split}
\end{equation}
where $N_s$ and $M_s$ are given, respectively, by 
\begin{subequations}
	\begin{equation}
	\begin{split}
	N_s=&\sinh^2(r_e)\cosh(2r_p)+\sinh^2(r_p)+(1/2)\sinh(2r_e)\sinh(2r_p)\cos(\theta_e+\theta_p),
	\end{split}
	\end{equation}
	\begin{equation}
	\begin{split}
	M_s=&\exp(\mathrm{i}\theta_p)\Big\{\frac{1}{2}\sinh(2r_p)\cosh(2r_e)+\frac{1}{2}\sinh(2r_e)\{\exp[\mathrm{i}(\theta_e+\theta_p)]\cosh^2(r_p)\\&+\exp[-\mathrm{i}(\theta_e+\theta_p)]\sinh^2(r_p)\}\Big\}.
	\end{split}
	\end{equation}
\end{subequations}
When we choose $r_e=r_p$ and $\theta_e+\theta_p=\pm n\pi\ (n=1, 3, 5 \cdots)$, $N_s$ and $M_s$ can be simplified to 0. In this way, the additional noise induced by squeezing the cavity mode is eliminated completely, and the master equation (\ref{Eq: master equ after squeezing}) is simplified to the standard Lindblad form 
\begin{equation}\label{Eq: final master equ}
\begin{split}
\dot{\rho}(t)=i[\rho(t),H'(t)]+\kappa_1\mathcal{L}[a_s]\rho(t)+\kappa_2\mathcal{L}[c]\rho(t)+\gamma\mathcal{L}[\sigma_-]\rho(t).
\end{split}
\end{equation}
In comparison with the case without exploiting the squeezing field (Figure \ref{fig2}b), the recovery of oscillations can be clearly observed, as shown in Figure \ref{fig2}c. In contrast1 with the case without squeezing the cavity mode (Figure \ref{fig2}a), a remarkable enhancement of atom-cavity coupling is observed in view of the shrink in oscillation period. The influence of the cavity decay $\kappa_1$ on the adiabatic elimination of the squeezed cavity mode $a_s$ is discussed in Appendix. The enhancement of atom-cavity coupling for different squeezing parameter $r_p$, defined as $g_{r_p}/g_0$ and obtained by comparing the oscillation periods for various $r_p$ with the period for $r_p=0$, is plotted in Figure \ref{fig2}d. It is noted that there is a small invalid region of enhancement with $g_{\mathrm{r_p}}/g_0<1$. This means that the threshold of the rise of the enhancement curve requires a larger squeezing parameter compared to previous report \cite{bib15}. This may be understood clearly from the analytical expression of the effective coupling (see Eq.(\ref{Eq:effective coupling})). Obviously, $g_{\mathrm{eff}}$ is approximately in direct proportion to two parts, i.e., exponential $r_p$ and inverse detuning difference. With the increase of $r_p$ in the small region, the enhancement of coupling contributed by exponential $r_p$ does not compensate the weakening induced by inverse detuning difference, resulting in a decrease in $g_{\mathrm{eff}}$ as a whole. For increasing $r_p$ outside the small region, the increase in exponential $r_p$ becomes dominated over the decrease in inverse detuning difference, therefore, the rise of $g_{\mathrm{eff}}$ curve can be observed. Nevertheless, the larger decay rates $\kappa_1=100g$, $\kappa_2=0.1g$, and $\gamma=0.1g$ can be taken for modest squeezing parameters, indicating the ability of our scheme to resist strong dissipations. Although in our scheme the auxiliary cavity is also required to possess a relatively high $Q$ (i.e., $\kappa_2$ < $g$), the rigorous restriction in cavity $Q$ (i.e., $g/\kappa_2 \sim 10^3$) can be loosened considerably via the cavity mode squeezing. Further, the squeezing can effectively enhance coupling strength $J$, so that ideal oscillations also occur for $J<g$ (not shown here). This could be of particular utility in real systems with weak coupling between neighbouring cavities.

\begin{figure}
	\centering
	\includegraphics[width=1\columnwidth]{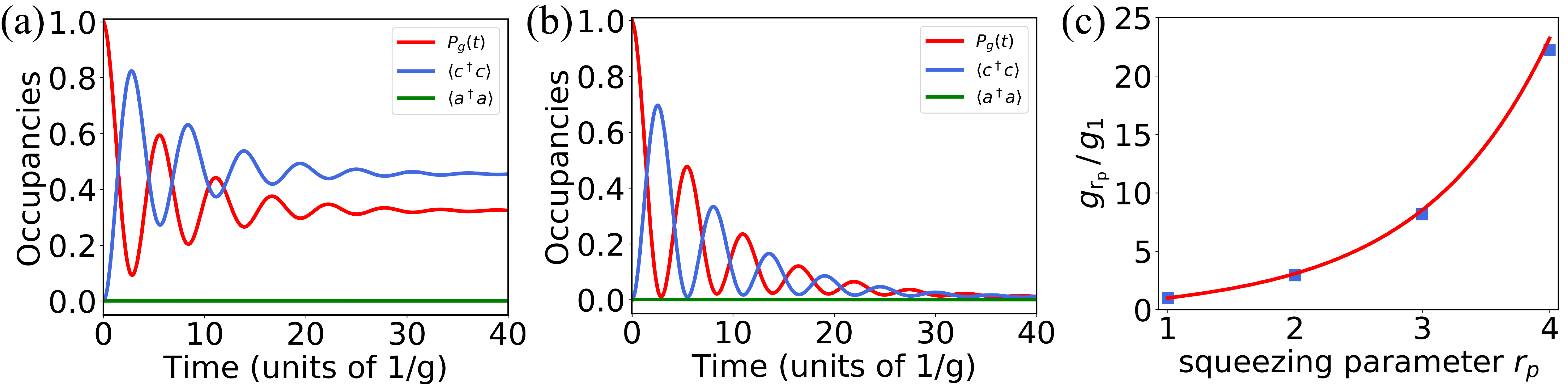}
	\caption{(a,b) Time evolution of the mean photon numbers $\langle a^\dagger a\rangle$ (green curves), $\langle c^\dagger c\rangle$ (blue curves), and the probability of the atom being in the ground state $P_g(t)$ (red curves). The parameters are: $r_p=r_e=4$, $\theta_p=0$, $\theta_e=\pi$, $\Delta_s+\Delta_q=25ge^{r_p}$; $\Delta_s-\Delta_q=20(\Delta_s+\Delta_q)$, $\kappa_1=100g$, (a) $\kappa_2=0.1g$, $\gamma=0.1g$, and (b)  $\kappa_2=0.2g$, $\gamma=10^{-3}g$. The initial states of the two cavities are in the vacuum states, and the atom is in the ground state. (c) Enhancement of atom-cavity coupling versus squeezing parameter $r_p$. The red curve is obtained from the analytical expression of the effective coupling, i.e., the values of $g_{\mathrm{eff}}'$ for arbitrary $r_p$ over the value of $g_{\mathrm{eff}}'$ for $r_p=1$; The blue squares are obtained by comparing the oscillation periods for various $r_p$ with the period for $r_p=1$.}
	\label{fig3}
\end{figure}

In what follows, we investigate the properties of the system when specific parameter conditions, $\Delta_q=-\Delta_c$ and $g=J$, are satisfied. Under the rotating-wave approximation condition ($\Delta_s-\Delta_q\gg\Delta_s+\Delta_q$) and large detuning condition [$\Delta_s+\Delta_q\gg(g/2)e^{r_p}$], the effective Hamiltonian between the atom and the auxiliary cavity mode $c$ is given by
\begin{equation}\label{Eq:effective H1}
	H_{\mathrm{eff}}^2=g_{\mathrm{eff}}'(\sigma_+c^\dag+c\sigma_-),
	\end{equation}
with the effective coupling coefficient being
\begin{equation}\label{Eq:effective couplings}
g_{\mathrm{eff}}'=\frac{gJ\cosh({r_p})\sinh({r_p})}{\Delta_s+\Delta_q}.
\end{equation}
By choosing specific parameter conditions given above, the rotating-wave interactions between the atom and the cavity mode $c$ can be eliminated effectively. We assume initially the atom is in the ground state and the two cavity modes are in their vacuum states. As shown in Figure \ref{fig3}a, oscillation between atomic ground state and cavity mode $c$ for several periods can be observed, with the maximum occupancy in mode $c$ exceeding 0.8. In the present system, the steady states of the atom and cavity mode $c$ are superposition states, which is different from that in the parameter regime in Figure \ref{fig2}c. When we further increase the difference between $\kappa_2$ and $\gamma$ (e.g., $\kappa_2=0.2g$ and $\gamma=10^{-3}g$), the dynamical evolution of the system is plotted in Figure \ref{fig3}b. As expected, the incease in cavity decay does not destroy the oscillations due to smaller atomic dissipation. In addition, we see that the steady state of the atom and the cavity mode $c$ become excited and vacuum states, respectively. When dissipations are not considered, the transition $\vert g\rangle \vert0\rangle_c \leftrightarrow \vert e\rangle \vert1\rangle_c$ occurs continually under the Hamiltonian (\ref{Eq:effective H1}). While a larger cavity decay compared to the atomic spontaneous emission will relax the state $\vert e\rangle \vert1\rangle_c$ to $\vert e\rangle \vert0\rangle_c$, resulting in the observed steady states of the system. Figure \ref{fig3}c shows the enhancement of coupling as a function of $r_p$ ranging from 1 to 4. For $r_p=0$, there is no coupling between the atom and the auxiliary cavity, i.e., $g_{\mathrm{eff}}'=0$. The red curve and blue squares correspond to analytical and numerical results, which are obtained from Eq. (\ref{Eq:effective couplings}) and the comparison in oscillation periods, respectively. With modest squeezing parameters, dramatically enhancement of atom-cavity coupling is achieved in the present parameter regime.

\section{Discussion and Conclusion}\label{sec4}
We briefly discuss the possible experimental implementations of the proposed scheme. The pumped $\chi^{(2)}$ nonlinear resonator (e.g., photonic crystal cavities \cite{bib34} or whispering gallery microcavities \cite{bib35}) coupled to an atomic emitter is a promising setup for realizing our scheme. The configuration of the two-level atom can be realized in alkali-metal atoms, e.g., cesium \cite{bib36} and rubidium \cite{bib4,bib37}. The high bandwidth squeezed field can be generated via pumping a second-order nonlinear medium, e.g., periodically-poled potassium titanyl phosphate (PPKTP) crystal \cite{bib24}. Squeezing of the cavity mode using PPKTP crystal is also demonstrated in \cite{bib38,bib39}. In addition, solid-state system can also be considered as an alternative implementation, particularly the circuit QED system where superconducting qubits are employed as two-level artificial atoms coupled with inductance/capacitance resonator or coplanar waveguide resonator \cite{bib40,bib41,bib42}. Squeezing inside the cavity is achievable by inserting a superconducting quantum interference device (SQUID) \cite{bib43,bib44}. Since the discussed model is generic, our scheme is not limited and could be applied to diverse physical systems.

In conclusion, we have demonstrated that parametrically squeezing the cavity mode enables enhancement of atom-field coupling in highly dissipative coupled cavities. By introducing squeezing into the primary cavity where the mode can be adiabatically eliminated for large detuning and adjusting the squeezing parameter, coupling enhancement as well as effective strong coupling between atom and auxiliary cavity are realized when specific conditions are satisfied. The additional noises of the squeezed mode can be suppressed via parametric matching with an auxiliary squeezed field. In comparison with existing schemes, our proposal allows larger rates of both atomic spontaneous emission and cavities decay due to the employment of squeezing. The restriction in cavity $Q$ as well as coupling strength between neighbouring cavities can be loosened considerably. Our method can be applicable to different physical system, and will find various applications in quantum information processing, e.g., entanglement preparation and quantum gate implementation.

\section*{Appendix}\label{appendix}
In this section, we mainly discuss the influence of the cavity decay $\kappa_1$ on the adiabatic elimination of the squeezed cavity mode $a_s$. In the interaction picture, the Hamiltonian (\ref{Eq:squeezing frame H}) can be divided into two parts in accordance with the rotating and counter-rotating-wave interactions 
\begin{subequations}\label{Eq:squeezing frame H1}
\begin{equation}\label{Eq: rotating interaction}
\begin{split}
H_{r}=g\cosh(r_p)a_s\sigma_+e^{\mathrm{i}(\Delta_q-\Delta_s)t}+J\cosh(r_p)a_sc^\dag e^{\mathrm{i}(\Delta_c-\Delta_s)t}+\mathrm{H.c.},
\end{split}
\end{equation}
\begin{equation}
\begin{split}
H_{cr}=g\sinh(r_p)a_s^\dag\sigma_+e^{\mathrm{i}(\Delta_q+\Delta_s)t}+J\sinh(r_p)a_s^\dag c^\dag e^{\mathrm{i}(\Delta_c+\Delta_s)t}+\mathrm{H.c.}.
\end{split}
\end{equation}
\end{subequations}
Considering that $\Delta_q$ is approximately equal to $\Delta_c$ and $\Delta_q+\Delta_s$ is much larger than $\Delta_q-\Delta_s$, there are two independent transiting channels, corresponding to detunings $\Delta_q(\Delta_c)-\Delta_s$ and $\Delta_q(\Delta_c)+\Delta_s$, respectively. We first consider the rotating interactions described by Hamiltonian (\ref{Eq: rotating interaction}). After performance of proper frame rotation, the Hamiltonian (\ref{Eq: rotating interaction}) can be approximately rewritten as 
\begin{equation}\label{Eq:r and cr}
\begin{split}
H'_{r}=g\cosh(r_p)(a_s\sigma_++\sigma_-a_s^\dag)+J\cosh(r_p)(a_sc^\dag+ca_s^\dag)-\Delta_r a_s^\dag a_s,
\end{split}
\end{equation}
where $\Delta_r=\Delta_q(\Delta_c)-\Delta_s$. Starting from Eq. (\ref{Eq:r and cr}), the standard Heisenberg-Langevin equation for operator $a_s$ is given by 
\begin{equation}\label{Langevin equation of a_s}
\begin{split}
\dot{a}_{s}=\mathrm{i}[H'_{r},a_s]-\kappa_1 a_s=\mathrm{2i}(-g\cosh(r_p)\sigma_--J\cosh(r_p)c+\Delta_ra_s)-\kappa_1a_s.
\end{split}
\end{equation}
On condition that the cavity mode $a_s$ is adiabatically eliminated, $\dot{a}_{s}=0$ should be satisfied. By solving Eq. (\ref{Langevin equation of a_s}), we obtain the effective operator form of $a_s$ corresponding to the rotating interactions
\begin{equation}\label{Eq:effective operator as}
a_{s,r}=\frac{g\cosh(r_p)\sigma_-+J\cosh(r_p)c}{\Delta_r+\mathrm{i}\kappa_1/2}.
\end{equation}
Following the same method as above, it is easy to obtain the effective operator form of $a_s$ corresponding to the counter-rotating interactions
\begin{equation}\label{Eq:effective operator as1}
a_{s,cr}=\frac{g\sinh(r_p)\sigma_++J\sinh(r_p)c^\dag}{-\Delta_{cr}+\mathrm{i}\kappa_1/2},
\end{equation}
where $\Delta_{cr}=\Delta_q(\Delta_c)+\Delta_s$. By substituting Eqs. (\ref{Eq:effective operator as}) and (\ref{Eq:effective operator as1}) to the standard Lindblad master equation, we obtain the effective master equation after eliminating the squeezed cavity mode
\begin{equation}\label{Eq: effective master equ}
\begin{split}
\dot{\rho}(t)=&i[\rho(t),H_{\mathrm{eff}}^1]+\kappa_2\mathcal{L}[c]\rho(t)+\gamma\mathcal{L}[\sigma_-]\rho(t)+\frac{\kappa_1}{\Delta_r^2+(\kappa_1/2)^2}\mathcal{L}[a'_{s,r}]\rho(t)\\&+\frac{\kappa_1}{\Delta_{cr}^2+(\kappa_1/2)^2}\mathcal{L}[a'_{s,cr}]\rho(t),
\end{split}
\end{equation}
where $a'_{s,r}$ and $a'_{s,cr}$ are given, respectively, by
\begin{subequations}
	\begin{equation}
	a'_{s,r}=g\cosh(r_p)\sigma_-+J\cosh(r_p)c,
	\end{equation}
	\begin{equation}
	a'_{s,cr}=g\sinh(r_p)\sigma_++J\sinh(r_p)c^\dag.
	\end{equation}
\end{subequations}

\begin{figure*}
	\centering
	\includegraphics[width=1\columnwidth]{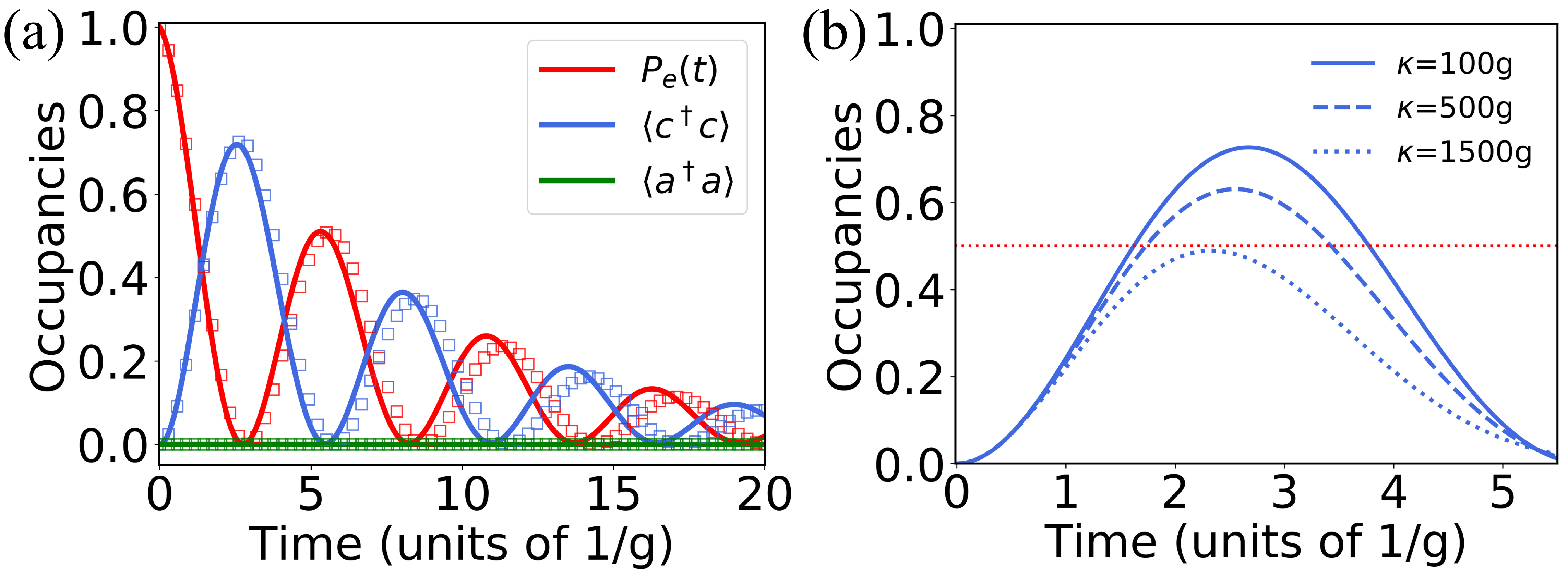}%
	\caption{(a) Time evolution of the mean photon numbers $\langle a^\dagger a\rangle$, $\langle c^\dagger c\rangle$ (blue curves), and the probability of the atom being in the excited state $P_e(t)$ (red curves). The curves and hollow squares are obtained by numerically solving the original and effective master equations, respectively. All parameters are the same as Figure \ref{fig2}c. (b) Time evolution of the mean photon number $\langle c^\dagger c\rangle$ in the first period for various $\kappa_1$ under the effective master equation. The horizontal line denotes the occupancy of 0.5.}
	\label{fig4}
\end{figure*}

Although the squeezed cavity mode with decay rate of $\kappa_1$ is eliminated, it is shown that the effective decay rates $\kappa_1/[\Delta_r^2+(\kappa_1/2)^2]$ and $\kappa_1/[\Delta_{cr}^2+(\kappa_1/2)^2]$ are functions of $\kappa_1$. Therefore, relatively large $\kappa_1$ could damp the effective interactions in the system after adiabatic elimination. To verify this, we first examine the validity of the effective master equation. Specifically, we repeat the plot shown in Figure \ref{fig2}c by numerically solving Eq. (\ref{Eq: effective master equ}), as shown in Figure \ref{fig4}a. It shows that the dynamical evolution of the system under the effective master equation agrees well with that under the original master equation, which exemplify the validity of the effective master equation. In Figure \ref{fig4}b, we plot the evolution of $\langle c^\dagger c\rangle$ for various $\kappa_1$ by numerically solving the effective master equation. The population of $\langle c^\dagger c\rangle$ drops below 0.5 when $\kappa_1$ increases to approximately $1500g$. With the further increasing of $\kappa_1$, the evolution of $\langle c^\dagger c\rangle$ may be suppressed completely, i.e., the adiabatic elimination as well as the resulting effective interaction may become invalid.

\section*{Acknowledgements}
This work was supported by the National Natural Science Foundation of China (NSFC) (11675046); Program for Innovation Research of Science in Harbin Institute of Technology (A201412); Postdoctoral Scientific Research Developmental Fund of Heilongjiang Province (LBH-Q15060).


\begin{thebibliography}{0}
	
\bibitem{bib1}% 
\textsc{Y.~Ota}, 
\textsc{R.~Ohta},
\textsc{N.~Kumagai}, 
\textsc{S.~Iwamoto}, and 
\textsc{Y.~Arakawa},
{\textit{Phys. Rev. Lett.}} \textbf{2015}, \textit{114}, 143603.

\bibitem{bib2}% 
\textsc{J.\,M.~Raimond}, 
\textsc{M.~Brune}, and 
\textsc{S.~Haroche},
{\textit{Rev.Mod.Phys.}} \textbf{2001}, \textit{73}, 565–582.

\bibitem{bib3}% 
\textsc{T.~Vogt}, 
\textsc{M.~Viteau},
\textsc{J.~Zhao}, 
\textsc{A.~Chotia},
\textsc{D.~Comparat}, and 
\textsc{P.~Pillet},
{\textit{Phys. Rev. Lett.}} \textbf{2006}, \textit{97}, 083003.

\bibitem{bib4}% 
\textsc{C.~Sames}, 
\textsc{H.~Chibani},
\textsc{C.~Hamsen}, 
\textsc{P.\,A.~Altin},
\textsc{T.~Wilk}, and 
\textsc{G.~Rempe},
{\textit{Phys. Rev. Lett.}} \textbf{2014}, \textit{112}, 043601.

\bibitem{bib5}% 
\textsc{A.~Wallraff}, 
\textsc{D.\,I.~Schuster},
\textsc{A.~Blais}, 
\textsc{L.~Frunzio},
\textsc{R.\,S.~Huang},
\textsc{J.~Majer},
\textsc{S.~Kumar},
\textsc{S.\,M.~Girvin}, and 
\textsc{R.\,J.~Schoelkopf},
{\textit{Nature}} \textbf{2004}, \textit{431}, 162.

\bibitem{bib6}% 
\textsc{A.~Blais}, 
\textsc{R.\,S.~Huang},
\textsc{A.~Wallraff}, 
\textsc{S.\,M.~Girvin}, and 
\textsc{R.\,J.~Schoelkopf},
{\textit{Phys. Rev. A}} \textbf{2004}, \textit{69}, 062320.

\bibitem{bib7}% 
\textsc{K.~Hennessy}, 
\textsc{A.~Badolato},
\textsc{M.~Winger}, 
\textsc{D.~Gerace},
\textsc{M.~Atat{\"u}re},
\textsc{S.~Gulde},
\textsc{S.~F{\"a}lt},
\textsc{E.\,L.~Hu}, and 
\textsc{A.~Imamo{\u{g}}lu},
{\textit{Nature}} \textbf{2007}, \textit{445}, 896.

\bibitem{bib8}% 
\textsc{C.~Jarlov}, 
\textsc{E.~Wodey},
\textsc{A.~Lyasota}, 
\textsc{M.~Calic},
\textsc{P.~Gallo},
\textsc{B.~Dwir},
\textsc{A.~Rudra}, and 
\textsc{E.~Kapon},
{\textit{Phys. Rev. Lett.}} \textbf{2016}, \textit{117}, 076801.

\bibitem{bib9}% 
\textsc{Y.~Ota}, 
\textsc{D.~Takamiya}, 
\textsc{R.~Ohta},
\textsc{H.~Takagi}, 
\textsc{N.~Kumagai},
\textsc{S.~Iwamoto}, and 
\textsc{Y.~Arakawa},
{\textit{Appl. Phys. Lett.}} \textbf{2018}, \textit{112}, 093101.

\bibitem{bib10}% 
\textsc{D.~Zueco}, and 
\textsc{J.~Garc\'{\i}a-Ripoll},
{\textit{Phys. Rev. A}} \textbf{2019}, \textit{99}, 013807.

\bibitem{bib11}% 
\textsc{J.~Ren}, 
\textsc{Y.~Gu}, 
\textsc{D.~Zhao},
\textsc{F.~Zhang}, 
\textsc{T.~Zhang}, and 
\textsc{Q.~Gong},
{\textit{Phys. Rev. Lett.}} \textbf{2017}, \textit{118}, 073604.

\bibitem{bib12}% 
\textsc{S.~Gr{\"o}blacher}, 
\textsc{K.~Hammerer}, 
\textsc{M.\,R.~Vanner}, and 
\textsc{M.~Aspelmeyer},
{\textit{Nature}} \textbf{2009}, \textit{460}, 724.

\bibitem{bib13}% 
\textsc{C.\,D.~Ogden}, 
\textsc{E.\,K.~Irish}, and 
\textsc{M.\,S.~Kim},
{\textit{Phys. Rev. A}} \textbf{2008}, \textit{78}, 063805.

\bibitem{bib14}% 
\textsc{Y.\,C.~Liu}, 
\textsc{X.~Luan}, 
\textsc{H.\,K.~Li},
\textsc{Q.~Gong},
\textsc{C.\,W.~Wong}, and 
\textsc{Y.\,F.~Xiao},
{\textit{Phys. Rev. Lett.}} \textbf{2014}, \textit{112}, 213602.

\bibitem{bib15}% 
\textsc{W.~Qin}, 
\textsc{A.~Miranowicz}, 
\textsc{P.\,B.~Li},
\textsc{X.\,Y.~L\"u},
\textsc{J.\,Q.~You}, and 
\textsc{F.~Nori},
{\textit{Phys. Rev. Lett.}} \textbf{2018}, \textit{120}, 093601.

\bibitem{bib16}% 
\textsc{C.~Leroux}, 
\textsc{L.\,C.\,G.~Govia}, and 
\textsc{A.\,A.~Clerk},
{\textit{Phys. Rev. Lett.}} \textbf{2018}, \textit{120}, 093602.

\bibitem{bib17}% 
\textsc{X.\,Y.~L\"u},
\textsc{Y.~Wu},
\textsc{J.\,R.~Johansson},
\textsc{H.~Jing},
\textsc{J.~Zhang}, and 
\textsc{F.~Nori},
{\textit{Phys. Rev. Lett.}} \textbf{2015}, \textit{114}, 093602.

\bibitem{bib18}% 
\textsc{Y.\,H.~Chen},
\textsc{W.~Qin}, and 
\textsc{F.~Nori},
{\textit{arXiv:1901.10249}} \textbf{2019}.

\bibitem{bib19}% 
\textsc{S.\,B.~Zheng}, and 
\textsc{G.\,C.~Guo},
{\textit{Phys. Rev. Lett.}} \textbf{2000}, \textit{85}, 2392–2395.

\bibitem{bib20}% 
\textsc{L.\,m.\,H.~S\'ark\'any},
\textsc{J.~Fort\'agh}, and 
\textsc{D.~Petrosyan},
{\textit{Phys. Rev. A}} \textbf{2018}, \textit{97}, 032341.

\bibitem{bib21}% 
\textsc{Y.~Wang},
\textsc{C.~Li},
\textsc{E.\,M.~Sampuli},
\textsc{J.~Song},
\textsc{Y.~Jiang}, and 
\textsc{Y.~Xia},
{\textit{Phys. Rev. A}} \textbf{2019}, \textit{99}, 023833.

\bibitem{bib22}% 
\textsc{C.~Qian}, 
\textsc{S.~Wu},
\textsc{F.~Song}, 
\textsc{K.~Peng},
\textsc{X.~Xie},
\textsc{J.~Yang},
\textsc{S.~Xiao},
\textsc{S.~Xiao},
\textsc{M.\,J.~Steer},
\textsc{I.\,G.~Thayne},
\textsc{C.~Tang},
\textsc{Z.~Zuo},
\textsc{K.~Jin},
\textsc{C.~Gu}, and 
\textsc{X.~Xu},
{\textit{Phys. Rev. Lett.}} \textbf{2018}, \textit{120}, 213901.

\bibitem{bib23}% 
\textsc{X.~Li}, 
\textsc{M.~Bamba},
\textsc{Q.~Zhang}, 
\textsc{S.~Fallahi},
\textsc{G.\,C.~Gardner},
\textsc{W.~Gao},
\textsc{M.~Lou},
\textsc{K.~Yoshioka},
\textsc{M.\,J.~Manfra}, and 
\textsc{J.~Kono},
{\textit{Nat. Photonics}} \textbf{2018}, \textit{12}, 324.

\bibitem{bib24}% 
\textsc{S.~Ast}, 
\textsc{M.~Mehmet}, and 
\textsc{R.~Schnabel},
{\textit{Opt. Express}} \textbf{2013}, \textit{12}, 13572–13579.

\bibitem{bib25}% 
\textsc{T.~Serikawa}, 
\textsc{J.~ichi\,Yoshikawa},
\textsc{K.~Makino}, and 
\textsc{A.~Frusawa},
{\textit{Opt. Express}} \textbf{2016}, \textit{24}, 28383–28391.

\bibitem{bib26}% 
\textsc{M.\,O.~Scully}, and 
\textsc{M.\,S.~Zubairy},
{\textit{Quantum optics}} Cambridge University Press, Cambridge, \textbf{1997}.

\bibitem{bib27}% 
\textsc{D.\,F.~James}, and 
\textsc{J.~Jerke},
{\textit{Can. J. Phys.}} \textbf{2007}, \textit{85}, 625–632.

\bibitem{bib28}% 
\textsc{H.\,P.~Breuer}, and 
\textsc{F.~Petruccione},
{\textit{The theory of open quantum systems}} Oxford University Press, New York, \textbf{2002}.

\bibitem{bib29}% 
\textsc{H.\,J.~Chen},
{\textit{Photon. Res.}} \textbf{2018}, \textit{6}, 1171–1176.

\bibitem{bib30}% 
\textsc{H.~Toida},
\textsc{T.~Nakajima}, and 
\textsc{S.~Komiyama},
{\textit{Phys. Rev. Lett.}} \textbf{2013}, \textit{110}, 066802.

\bibitem{bib31}% 
\textsc{A.~Ask},
\textsc{M.~ Ekstr\"om},
\textsc{P.~ Delsing}, and 
\textsc{G.~Johansson},
{\textit{Phys. Rev. A}} \textbf{2019}, \textit{99}, 013840.

\bibitem{bib32}% 
\textsc{H.\,J.~Carmichael},
\textsc{R.\,J.~Brecha},
\textsc{M.\,G.~ Raizen},
\textsc{H.\,J.~ Kimble}, and 
\textsc{P.\,R.~Rice},
{\textit{Phys. Rev. A}} \textbf{1989}, \textit{40}, 5516–5519.

\bibitem{bib33}% 
\textsc{J.~Johansson},
\textsc{P.~Nation}, and 
\textsc{F.~Nori},
{\textit{Comput. Phys. Commun.}} \textbf{2013}, \textit{184}, 1234–1240.

\bibitem{bib34}% 
\textsc{W.\,T.\,M.~Irvine},
\textsc{K.~Hennessy}, and 
\textsc{D.~Bouwmeester},
{\textit{Phys. Rev. Lett.}} \textbf{2006}, \textit{96}, 057405.

\bibitem{bib35}% 
\textsc{J.\,U.~F\"urst},
\textsc{D.\,V.~Strekalov},
\textsc{D.~Elser},
\textsc{A.~Aiello},
\textsc{U.\,L.~Andersen},
\textsc{C.~Marquardt}, and 
\textsc{G.~Leuchs},
{\textit{Phys. Rev. Lett.}} \textbf{2011}, \textit{106}, 113901.

\bibitem{bib36}% 
\textsc{A.~Boca},
\textsc{R.~Miller},
\textsc{K.\,M.~Birnbaum},
\textsc{A.\,D.~Boozer},
\textsc{J.~McKeever}, and 
\textsc{H.\,J.~Kimble},
{\textit{Phys. Rev. Lett.}} \textbf{2004}, \textit{93}, 233603.

\bibitem{bib37}% 
\textsc{P.~Maunz},
\textsc{T.~Puppe},
\textsc{I.~Schuster},
\textsc{N.~Syassen},
\textsc{P.\,W.\,H.~Pinkse}, and 
\textsc{G.~Rempe},
{\textit{Phys. Rev. Lett.}} \textbf{2005}, \textit{94}, 033002.

\bibitem{bib38}% 
\textsc{R.~Schnabel},
{\textit{Phy. Rep.}} \textbf{2017}, \textit{684}, 1–51.

\bibitem{bib39}% 
\textsc{H.~Vahlbruch},
\textsc{M.~Mehmet},
\textsc{K.~Danzmann}, and 
\textsc{R.~Schnabel},
{\textit{Phys. Rev. Lett.}} \textbf{2016}, \textit{117}, 110801.

\bibitem{bib40}% 
\textsc{I.~Chiorescu},
\textsc{P.~Bertet},
\textsc{K.~Semba},
\textsc{Y.~Nakamura},
\textsc{C.~Harmans}, and 
\textsc{J.~Mooij},
{\textit{Nature}} \textbf{2004}, \textit{431}, 159.

\bibitem{bib41}% 
\textsc{J.~You}, and
\textsc{F.~Nori},
{\textit{Nature}} \textbf{2011}, \textit{474}, 589.

\bibitem{bib42}% 
\textsc{S.\,J.~Xiong},
\textsc{Z.~Sun},
\textsc{J.\,M.~Liu},
\textsc{T.~Liu}, and 
\textsc{C.\,P.~Yang},
{\textit{Opt. Lett.}} \textbf{2015}, \textit{40}, 2221–2224.

\bibitem{bib43}% 
\textsc{K.~Moon},and 
\textsc{S.\,M.~Girvin},
{\textit{Phys. Rev. Lett.}} \textbf{2005}, \textit{95}, 140504.

\bibitem{bib44}% 
\textsc{S.~Kono},
\textsc{Y.~Masuyama},
\textsc{T.~Ishikawa},
\textsc{Y.~Yamazaki},
\textsc{K.~Usami},
\textsc{K.~Koshino}, and 
\textsc{Y.~Nakamura},
{\textit{Phys. Rev. Lett.}} \textbf{2017}, \textit{119}, 023602.
\end{thebibliography}
\end{document}